# Structural characterisation of tin fluorophosphate glasses doped with $Er_2O_3$

J. Trimble,[1] R. Golovchak,[1] J. Oelgoetz,[1] C. Brennan[2] & A. Kovalskiy[1]*

[1] Department of Physics and Astronomy, Austin Peay State University, Clarksville, TN 37044, USA
[2] Department of Chemistry, Austin Peay State University, Clarksville, TN 37044, USA

*EXAFS and confocal Raman microscopy have been used to study $50SnF_2.(20-x)SnO.30P_2O_5.xEr_2O_3$ (x=0, 0·1, 0·25) glasses. EXAFS data reveal an average coordination of Sn to O of 1·5 in both undoped glass and Er-doped glass samples. The first coordination sphere of Er in glasses doped with $Er_2O_3$ was found to have 9 F atom neighbours at an average bond length of 2·292±0·005 Å, indicating that Er preferentially bonds with F. Raman spectra clearly show the emergence of orthophosphate $Q^0$ units on addition of $Er_2O_3$. The saturation of Er solubility is found to be between 0·25 and 0·5 mol% $Er_2O_3$. An increase of the glass transition temperature from ~80°C in undoped glass to ~87°C in samples doped with 0·25 mol% $Er_2O_3$ was also observed. This process is accompanied by an increase in the difference between the crystallisation and glass transition temperatures, which is usually associated with improved thermal stability of the glass.*

## 1. Introduction

Fluorophosphate glasses have been studied as promising media for many applications, such as optical amplifiers, upconverting optical devices[1,2] and low temperature sealing agents.[3] Compositions of these glasses have also been developed as a host matrix for organic dopants for use in optical signal processing devices.[4] However, these materials have exhibited poor chemical durability, which has limited their technological application. The addition of rare earth elements or other dopants can improve the chemical stability of the glass,[1,5] as well as introducing additional functionality for active optical applications.[1] A variety of network modifying cations may be added to fluorophosphate glass compositions to engineer their properties.[1–3,5,6] One such element is Sn, which lowers the melting temperature of fluorophosphate glasses. The most durable compositions among tin fluorophosphate glasses have been found to have ~30 mol% $P_2O_5$; this is because less than 5% of O in these glasses is involved in P–O–P bridging oxygen bonds.[5] These bonds are probably more susceptible to reacting with water molecules from the air.[5]

The atomic structure of these glasses has been studied previously using Raman, XPS and infrared spectroscopy.[5,7–9] Phosphorus cations are believed to be four-fold coordinated and primarily exist in tetrahedral geometry as $FPO_3$ monomers or $P_2O_7$ dimers in glasses with 30 mol% $P_2O_5$.[5,9] Sn is expected to be three-fold[7] or four-fold[10] coordinated, in trigonal pyramidal or tetrahedral geometry.

It was previously shown that the addition of $Er_2O_3$ to the matrix of phosphate and fluorophosphate glasses strengthens network connectivity and improves their chemical durability.[1,11,12] Also, the addition of $Er_2O_3$ or $ErF_3$ significantly contributes to the efficiency of energy conversion processes in fluorophosphate glasses.[1,13] However, the mechanism of Er incorporation, the specifics of structural organisation, and the dependence on Er concentration, are not fully clear, especially in the tin fluorophosphate matrix.

The atomic or molecular structure of Er-doped fluorophosphate glass is still unclear, despite previous studies.[6,14] In the present study, advanced experimental methods, such as EXAFS (extended x-ray absorption fine structure) and Raman microscopy, were used to investigate the structure of undoped and Er-doped tin fluorophosphate glasses. The main goal of this work is to clarify the coordination of Sn and Er in the glass matrix, and to verify the molecular structure of the glasses, especially in Er-doped compositions.

## 2. Experimental procedure

Samples of $50SnF_2.(20-x)SnO.30P_2O_5.xEr_2O_3$ (x=0, 0·1, 0·25) glass have been investigated. Twelve gram melts of reagent grade $SnF_2$, SnO, $P_2O_5$ and $Er_2O_3$ were mixed in an argon atmosphere and melted in alumina crucibles at 330°C for 2 h, and 400°C for an additional hour in air. The samples were stirred after 2 h and again 15 min before being cast. Glass disks of ~2 cm diameter and 0·5 cm thickness were quenched onto copper plates and stored in desiccators.

The Sn:P and O:F ratios were obtained from energy dispersive x-ray spectroscopy using a Hitachi TM-

1000 scanning electron microscope. The experimentally determined Sn:P ratios in the final composition were ~10% higher in the undoped glass and ~1–5% lower in doped glass than the nominal ratio (~1·16). The O:F ratio in the final glass compositions was found to be ~10% higher (undoped), or ~15% higher (doped samples) than the nominal composition ratio (~1·7), indicating the loss of F during synthesis probably through the evolution of HF gas. Aluminium impurities from the crucible were detected at ~1 wt%.

Undoped samples and samples doped with 0·1 mol% $Er_2O_3$ were transparent and colourless, while the samples doped with 0·25 mol% $Er_2O_3$ have a faint pink tint. Data from samples doped with 0·5 mol% $Er_2O_3$ and higher, which are a translucent pink with visible phase separation, were excluded from consideration. All measurements were made before any corrosion became noticeable. Within one week, a strong corrosion was observed in the undoped samples exposed to a normal atmosphere, with a white surface film covering the glass. The effect of this corrosion is significantly diminished with increasing $Er_2O_3$ concentration.

The EXAFS measurements were performed at the X18B x-ray beamline at the National Synchrotron Light Source at Brookhaven National Laboratory. The samples were powdered and glued onto Kapton tape, and data were collected at the Sn K-edge (29·2 KeV) and the Er $L_3$-edge (8·358 KeV) in fluorescent mode, using a Passivated Implanted Planar Silicon (PIPS) detector.

In general, the EXAFS signal represents modulation in the absorption coefficient, $\mu_i$, as a function of x-ray energy, $E=\hbar\omega$. Using the one-electron approximation of Fermi's Golden Rule for $\mu_i(\omega)$, under the plane wave approximation, the EXAFS equation for isotropic materials can be expressed as[15]

$$\mu_i(k) = \frac{\mu_1(k)}{\mu_0(k)} = \sum_j \frac{n_j S_{0j}^2(k) F_j(k) \sin(2kR_j + \varphi_j(k))}{kR_j^2 \exp\left(\frac{2R_j}{\Lambda} + 2k^2\sigma_{0j}^2\right)} \quad (1)$$

where $\mu_1(k)$ and $\mu_0(k)$ are the terms contributed by single scattering and background, respectively (multiple scattering terms are negligible for the EXAFS region). $k \approx 0.512(E-E_0)^{1/2}$ Å$^{-1}$ is the photoelectron wavevector, $E_0$ is the threshold energy, $S_{0j}$ is the passive electron reduction factor, $n_j$ is the degeneracy of the path, $N_j = nS_{0j}^2(k)$ is the number of neighbours in the $j$th shell at an average distance $R_j$, $F_j(k)$ is the effective amplitude of the backscattered electron wave, $\phi_j(k)$ is the effective phase shift between backscattered and outgoing electron wave, $\Lambda$ is the mean free path of photoelectrons, and $\sigma_{0j}^2$ is the the Debye–Waller factor related with disorder.

The standard Athena-Artemis software[16] package was used to process the experimental EXAFS spectra. A Kaiser-Bessel type of window function[17] was applied for restricting the EXAFS data in $k$-space. The Artemis software was used to generate an input file for FEFF calculations,[16] using crystallographic data for $Sn_3(PO_4)_2$[19] and $ErF_3$.[20] The calculated scattering paths for the nearest neighbours were used to fit the first shell in $k$-space of Sn and Er, exploiting the Levenberg–Marquardt method of nonlinear least-squares minimisation[18] implemented in Artemis (the energy shift, $\Delta E_0$, was fixed during the fitting).

Raman spectroscopy measurements were taken with an Xplora (Horiba Jobin-Yvon) Raman confocal microscope using 785 nm 52 mW excitation and an 1800 groove grating. 150 acquisitions of 12 s exposures using a 300 μm hole and a 200 μm slit were averaged. Scans with spectral resolution of ~3·5 cm$^{-1}$ were performed at room temperature.

Differential scanning calorimetry (DSC) measurements were made with a TA Instruments 2920 differential scanning calorimeter. A heating rate of 5°C/min was used to scan samples of ~22 mg in a nitrogen atmosphere. The glass transition temperature $T_g$ was taken as the onset value of the endothermic shift of the baseline, whereas the crystallisation temperature $T_x$ was determined from the peak value of the corresponding exotherm. Based on previous studies using this instrument, we estimate the experimental uncertainty of the reported $T_g$ and $T_x$ values as ±2°C.

## 3. Results and discussion

Raman spectra of the $50SnF_2.(20-x)SnO.30P_2O_5.xEr_2O_3$ ($x=0, 0.1, 0.25$) glasses are shown in Figure 1. Peaks in the high frequency region of the spectrum have previously[5,9,21] been identified as follows: the peak at 1035 cm$^{-1}$ is assigned to the P–O symmetric stretching vibration of the pyrophosphate $Q^1$ unit[5,21] (where $Q^n$ denotes a $PO_4$ tetrahedron with $n$ bridging oxygen atoms) or the fluorophosphate monomer, $FPO_3$.[5,9] The band near 860 cm$^{-1}$ is associated with the P–F symmetric stretch,[5] and the band near 740 cm$^{-1}$ is related to a P–O–P symmetric stretching mode.[5,21]

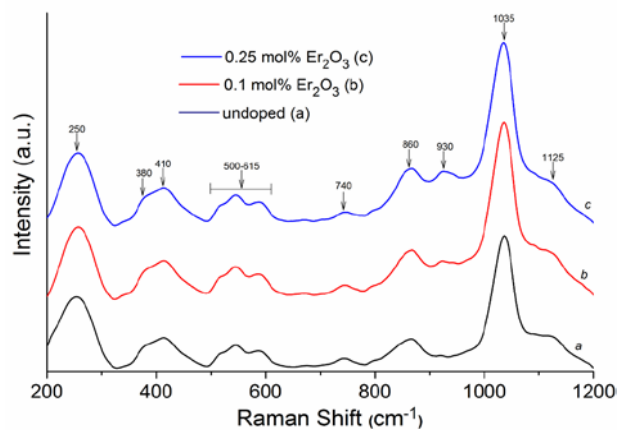

*Figure 1. Raman spectra of $50SnF_2.(20-x)SnO.30P_2O_5.xEr_2O_3$ ($x=0, 0.1, 0.25$) glasses [Colour available online]*

The peak near 1125 cm$^{-1}$ is associated with the P–O symmetric stretch of the metaphosphate $Q^2$ unit,[21] which exists residually in our undoped glass. In the low frequency region of the spectrum, the band near 250 cm$^{-1}$ has been associated with Sn–O$_2$ bending modes.[22] The bands between 500–615 cm$^{-1}$ have not been described in detail but this region has previously been associated with Sn and P vibrations without specific assignment.[5,23,24] Based on our measurements of the spectrum of crystalline SnF$_2$ (not shown), we assign the peaks near 380 and 410 cm$^{-1}$ to Sn–F fragments. This assignment is in good agreement with theoretical calculations of Sn–F vibrations.[25]

The effect of the Er dopant on the Raman spectra of the tin fluorophosphate glass is shown in Figure 1(b) and (c). The band near 930 cm$^{-1}$ that emerges with Er$_2$O$_3$ addition is assigned to the P–O symmetric stretch of isolated orthophosphate $Q^0$ tetrahedra.[5] There is also a slight decrease in intensity of the band near 1125 cm$^{-1}$ that is associated with $Q^2$ units. From the experimental spectra we cannot make trustworthy conclusions about the changes of the peak near 1035 cm$^{-1}$ that is associated with FPO$_3$ or $Q^1$ units, or the band near 740 cm$^{-1}$, but nevertheless we can note the tendency of the former peak to decrease and the latter to increase. These changes in the Raman spectra are consistent with the depolymerisation of the phosphate chain as the [P]/[O] ratio decreases when Er$_2$O$_3$ replaces SnO. The P–F band near 860 cm$^{-1}$ narrows with Er addition, but does not change significantly in intensity. No bands are observed in any of the measured Raman spectra in the region from 1125 to 1500 cm$^{-1}$.

Our EXAFS studies allow the determination of average coordination numbers, bond lengths and Debye-Waller factors for both Sn and Er atoms (Table 2, Figure 2). The first coordination shell of Sn was fitted with the Sn→O path only, since including the Sn→F path, derived from the known crystalline structures, did not lead to convergence of the fit. The possible reason could be that Sn–F interatomic distances in known crystal structures differ too much from those in the glass structure, or that the EXAFS oscillations corresponding to the Sn→F path in the investigated glasses significantly overlap with the background and, thus, cannot be extracted with confidence. Another explanation could be that the Sn–F bond lengths in the glass are very irregular (covering a broad range), which prevents a steady pattern in the modulation of the absorption coefficient (and, thus, the related EXAFS oscillations to occur) as is the case for the second and further coordination spheres in disordered solids (no peaks corresponding to these coordination spheres could be observed in the partial radial distribution function). So, it is possible that the observed spectrum is dominated by the EXAFS oscillations of Sn→O path, where the Sn–O bond length is relatively well defined throughout the structure. The obtained Sn–O bond length is 2·114±0·005 Å in the undoped sample, increasing to 2·119 and 2·121±0·005 Å in the samples doped with 0·1 and 0·25 mol% Er$_2$O$_3$, respectively. The Sn–O coordination is found to be 1·5 in the undoped sample, and remains constant (within the experimental uncertainty) in the two doped samples. If we assume the total coordination for Sn to be three, we can speculate that Sn is bonded to 1–2 oxygen atoms and the remaining bonds are with F. This is in good agreement with the structural model proposed by Brow et al.[7] Missing paths could be a reason for the slight mismatch of the experimental data and the fitting at higher $k$ values (Figure 2(b)) in the Sn K-edge EXAFS spectra. The Er L$_3$-edge EXAFS spectra are noisier because of the low concentration of Er in the investigated samples. Nevertheless, it was possible to obtain an acceptable goodness of fit with an Er coordination of approximately nine in the sample doped with 0·25 mol% Er$_2$O$_3$ and an Er–F bond length of 2·292±0·005 Å. This Er–F bond length is slightly greater than the value 2·280 Å, as measured by EXAFS for Er–F bonds in crystalline $ErF_3$.[26] Note that each Er in crystalline $ErF_3$ has nine first neighbour F atoms at six different lengths in the range of 2·248–2·589 Å.[20]

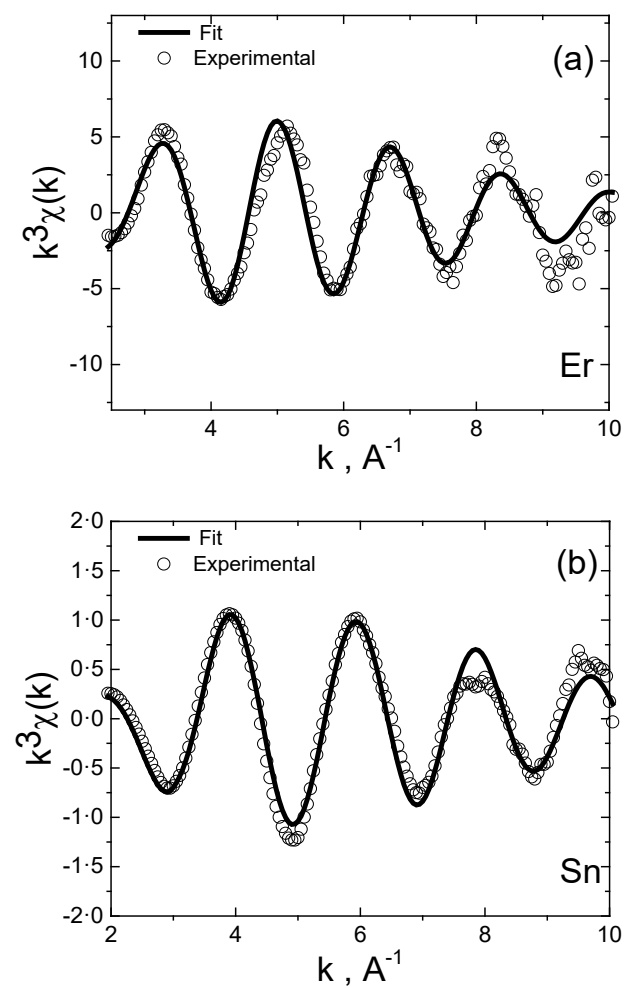

Figure 2. Examples of $k^3$-weighted EXAFS oscillations (symbols) and fits (solid lines) for (a) the Er L$_3$-edge of the 50SnF$_2$.19·75SnO.30P$_2$O$_5$.0·25Er$_2$O$_3$ glass, and (b) the Sn K-edge of the undoped 50SnF$_2$.(20−x)SnO.30P$_2$O$_5$ glass

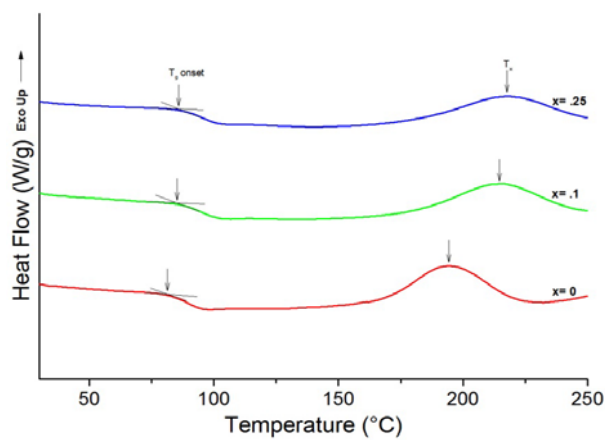

*Figure 3. DSC thermograms of $50SnF_2.(20-x)SnO.30P_2O_5.xEr_2O_3$ (x=0, 0·1, 0·25) glasses, heated at 5°C/min [Colour available online]*

DSC thermograms of the glass samples are shown in Figure 3, and the measured values for $T_g$, $T_x$, and $T_x$–$T_g$, are given in Table 1. The glass transition temperature for the undoped sample of 80°C increases by 5–7°C with Er addition. The Dietzel thermal stability criterion[27] (calculated as $T_x$–$T_g$) improves with Er doping. It should be noted that the value $T_g$=95°C previously reported for the undoped glass[8] (measured at 4°C/min) was obtained for samples with a different thermal history (quenching temperature of 500°C versus 400°C). Differences in $T_g$ values such as this have been reported for fluorophosphate glasses before, and have been attributed to variations in F content in the final glass composition.[28] The increase of $T_g$ and $T_x$ in the doped samples indicates that the addition of $Er_2O_3$ strengthens the glass network.[1,12] Stronger ionic bonds between $Er^{3+}$ and neighbouring anions replace weaker bonds to $Sn^{2+}$, which improves the thermal properties. The increase in characteristic temperatures may also be partially attributed to the greater density of network linkages that occurs when a nine coordinated Er–F environment replaces mixed three coordinated Sn–[F,O] bonding.

The structural model of Brow et al[7] is based on XPS analysis of tin fluorophosphate glass with high $SnF_2$ content and a ratio $[SnO+SnF_2]:P_2O_5>1$. This model is valid for the glass composition studied in the present paper. In this model, phosphorus cations are four-fold coordinated and exist in tetrahedral geometry as $FPO_3$. Sn is three-fold coordinated and is at the apex of a trigonal pyramid with a lone pair of electrons. Sn is linked to phosphorus via bridging oxygen, and is linked to another Sn via a bridging F. The third neighbour of Sn in this model alternates between O and F, which on average gives a Sn–O coordination of 1·5, in good agreement with our data. This finding is supported by previous research that found both P–F and Sn–F bonds are features of these glasses.[8,29] The presence of $FPO_3$ units for phosphate glass compositions with ~30 mol% $P_2O_5$ has been well studied.[5,9] Aside from this work, we know of no EXAFS data for tin fluorophosphate glasses that give experimental confirmation of the Sn coordination sphere. Thus the results presented in this paper confirm the Sn–O coordination in the structural model given by Brow et al.[7]

The coordination number of ~9 for Er atoms in $50SnF_2.19·75SnO.30P_2O_5.0·25Er_2O_3$ glass is consistent with the $ErF_3$ crystal structure (orthorhombic cell, space group *pnma*[20]). In this crystal, Er has a total of 13 neighbours, with nine closest neighbours followed by two more at a slightly longer bond length, and another two more at an even longer bond length. This result suggests that $Er_2O_3$ acts as a network modifier in our glasses, and that F migrates to Er sites to preserve charge neutrality, which would be consistent with the formation of $Q^0$ fragments observed in the Raman spectra. The preference for F coordination to the rare earth in tin fluorophosphate glass may result from the formation of P–F bonds reducing the polarisability of the corresponding P–O bonds, which typically fill the coordination of rare earths with O.[6,30] Preferential covalency of Er bonding with O in fluorophosphate glass has been reported for Al-containing compositions[6,30] where only a few P–F bonds form. In tin fluorophosphate glass, Sn–F bonds are not preferred (only forming in a limited number to prevent Sn–O–Sn linkages[5]), whereas, for example, in Al–F–P–O glass strong Al–F bonds form.[31] The affinity of the cation modifier to bond with F in fluorophosphate glass determines the extent of P–F bond formation, which in turn determines the preference of coordination of rare earths to O or F. It should be noted that Er may still be involved in interchain bonding in structures such as Sn–F–Er–F–Sn, while Er–F terminal fragments are also likely to be present.

## 4. Conclusions

It has been found that $Er_2O_3$ can replace SnO in $50SnF_2.30P_2O_5.20SnO$ glasses up to 0·25 mol% without the formation of a crystalline phase. EXAFS and Raman data support the hypothesis that Er bonds principally with F in the glass, with an Er coordination number of ~9 in the doped glass samples, as determined by EXAFS, consistent with the $ErF_3$ structure. Preferential Er-F bonding in tin fluorophosphate glass may be attributed to the presence of P–F bonds, lowering the polarisability of the nonbridging P–O bonds, which would otherwise fill the coordination of Er with O. The EXAFS data also suggest that the coordination number of Sn to O is close to 1·5, and remains constant in doped samples within the uncertainty of the method. The Raman spectra confirm the formation of isolated $Q^0$ orthophosphate units on addition of $Er_2O_3$. It was confirmed that $T_g$ increases with the addition of Er in doped glass samples, and that their thermal stability improves due to an increase in the value of $T_x$–$T_g$ (according to the Dietzel criterion).

*Table 1. $T_g$, $T_x$ and $T_x$–$T_g$ values for the doped and undoped tin fluorophosphate glasses*

| Batch Composition (mol%) | Appearance | $T_g$ (°C) | $T_x$ (°C) | $T_x$–$T_g$ (°C) |
| --- | --- | --- | --- | --- |
| $50SnF_2.30P_2O_5.20SnO$ | Transparent, colourless | 80 | 194 | 114 |
| $50SnF_2.30P_2O_5.19·9SnO.0·1Er_2O_3$ | Transparent, colourless | 85 | 216 | 131 |
| $50SnF_2.30P_2O_5.19·75SnO.0·25Er_2O_3$ | Transparent, light pink | 87 | 218 | 131 |

*Table 2. Bond lengths, Debye–Waller factors, and coordination numbers of Sn and Er in undoped and doped $50SnF_2.30P_2O_5.20SnO$ glasses*

| Composition | Edge | | | | | |
|---|---|---|---|---|---|---|
| | Er $L_3$ | | | Sn K | | |
| | $r$ (Å) (±0·005) | $\sigma_0$ (Å$^2$) (±0·0014) | $N_{Er}$ (±1·3) | $r$ (Å) (±0·005) | $\sigma_0$ (Å$^2$) (±0·0012) | $N_{Sn}$ (±0·10) |
| Undoped | | | | 2·114 | 0·0074 | 1·50 |
| Doped with 0·1 mol% $Er_2O_3$ | | | | 2·119 | 0·0082 | 1·55 |
| Doped with 0·25 mol% $Er_2O_3$ | 2·292 | 0·0103 | 9·0 | 2·121 | 0·0083 | 1·59 |

## Acknowledgements

We would like to thank the NSF International Materials Institute for New Functionality in Glass (NSF Grant No. DMR 0844014) and the NASA Tennessee Space Grant Consortium for funding. The Horiba Fluorolog 3 and Horiba Confocal Raman Microscope were purchased under NASA grant/cooperative agreement NNX10AJ04G. EXAFS data were taken at the X18B NSLS beamline at Brookhaven National Laboratory. The scientific and technical support of Dr. Syed Khalid (NSLS), and Chatree Saiyasombat (Lehigh University) are highly appreciated. The authors would like also to acknowledge the intellectual contribution of former undergraduate students at Austin Peay State University: Nichole Boyer, Tristan Harper and James York-Winegar.